\documentclass[aps,pra,amsmath,amssymb,reprint,superscriptaddress]{revtex4-2}

\usepackage[colorlinks,citecolor=blue,linkcolor=red,urlcolor=blue]{hyperref}
\usepackage{color}
\usepackage{graphicx}
\usepackage{subfigure}
\usepackage{verbatim}
\usepackage[normalem]{ulem}
\usepackage{bbm}
\usepackage{bbold} 

\begin{document}

\preprint{APS/123-QED}

\title{Tunable non-additivity in Casimir-Lifshitz force between graphene gratings}

\affiliation{Laboratoire Charles Coulomb (L2C), UMR 5221 CNRS-{Universit\'e de} Montpellier, F-34095 Montpellier, France}
\author{Youssef Jeyar}
\email{youssef.jeyar@umontpellier.fr}
\author{Minggang Luo}
\affiliation{Laboratoire Charles Coulomb (L2C), UMR 5221 CNRS-{Universit\'e de} Montpellier, F-34095 Montpellier, France}
\author{Kevin Austry}
\affiliation{Laboratoire Charles Coulomb (L2C), UMR 5221 CNRS-{Universit\'e de} Montpellier, F-34095 Montpellier, France}
\author{Brahim Guizal}
\affiliation{Laboratoire Charles Coulomb (L2C), UMR 5221 CNRS-{Universit\'e de} Montpellier, F-34095 Montpellier, France}
\author{Yi Zheng}
\affiliation{Department of Mechanical and Industrial Engineering, Northeastern University, Boston, MA, 02115, USA}
\affiliation{Department of Chemical Engineering,  Northeastern University,  Boston, MA, 02115, USA}
\author{H. B. Chan}
\affiliation{Department of Physics, The Hong Kong University of Science and Technology, Clear Water Bay, Kowloon, Hong Kong, China}
\affiliation{William Mong Institute of
Nano Science and Technology, The Hong Kong University of Science and Technology, Clear Water Bay, Kowloon, Hong Kong, China}
\affiliation{Center for
Metamaterial Research, The Hong Kong University of Science and Technology, Clear Water Bay, Kowloon, Hong Kong, China}
\author{Mauro Antezza}
 \email{mauro.antezza@umontpellier.fr}
\affiliation{Laboratoire Charles Coulomb (L2C), UMR 5221 CNRS-{Universit\'e de} Montpellier, F-34095 Montpellier, France}
\affiliation{Institut Universitaire de France, 1 rue Descartes, Paris Cedex 05, F-75231, France}

\date{\today}

\begin{abstract}

We investigate the Casimir-Lifshitz force (CLF)  {between two identical graphene strip gratings, laid on finite dielectric substrates, by using the scattering matrix (S-matrix) approach derived from the Fourier Modal Method with Local Basis Functions (FMM-LBF)}. We fully take into account the high-order electromagnetic diffractions, the multiple scattering and the exact 2D feature of the graphene strips. We show that the non-additivity, which is one of the most interesting features of the CLF in general, is significantly high and can be  modulated {\it{in situ}}, without any change in the actual material geometry and this by {varying} the graphene chemical potential. {We discuss the nature of the geometrical effects and show the relevance of the geometric parameter $d/D$ (i.e. the ratio between separation and grating period), which allows to explore the regions of parameters where the additive result is fully acceptable or where the full calculation is needed.} This study can open to deeper experimental exploration of the non-additive features of the CLF with micro- or nano-electromechanical graphene-based systems.
\end{abstract}

\maketitle


\section{\label{sec:Intro}Introduction}

The Casimir-Lifshitz force (CLF) exists between any couple of electrically neutral bodies, and is due to both vacuum and thermal  electromagnetic field fluctuations. It has been widely investigated from both theoretical and experimental sides and for different  geometrical configurations, e.g., plane-plane \cite{Bressi2002prl}, sphere/particle-plane \cite{Lamoreaux1997prl,Mohideen1998prl,Decca2003prl,Antoine2009prl,Torricelli2010pra,Bimonte2017EL}, sphere-sphere \cite{Garrett2018prl,Nunes2021universe,Bimonte2018prd}, grating-grating \citep{Noto2014pra_dielectirc,Lambrecht2008prl,Wang2021NC,Intravaia2012pra,Mauro2020prl}, and sphere-grating \cite{Chan2008prl,Messina2015pra,Contreras-Reyes2010pra,Decca2013NC}, to name a few. {In particular,} gratings {lead to the excitation of} high-order diffraction modes {that play} a relevant role in the CLF.  { Additionally, the dielectric polarizability of the interacting bodies plays a crucial role in determining the magnitude and characteristics of this force.}

For a system with a complex structure (e.g., gratings), different parts interact with each other, which results in a complicate{d} calculation for its Casimir interaction \cite{Wang2021NC}. One of the features that makes the CLF interesting, and hard to compute at the same time, is that it is {inherently} a {\it non-additive} phenomenon. More specifically, the force acting on an object cannot be computed as a simple sum over the force that would act individually on the different components of the object itself. Due to the complex light-matter interaction, the fluctuations of the constituent electric dipoles are affected by the presence of other fluctuating dipoles in the structure, and a full collective analysis needs to be done to tackle these non-additive CLF effects \cite{Woods2016rmp,Chan2008prl,Rauno2005prl,Emig2001prl,Wang2021NC}.

Recently it has been shown that planar graphene structures {exhibit novel behaviors} in CLFs  in and out of thermal equilibrium \cite{PhysRevB.80.245424,Pablo2022tdm,Liu2021prl,Liu2021prb,Chahine2017prl,jeyarprb2023,D3CP03706A,wang2023photon,rodriguez2023giant}, as well as in radiative heat transfer \cite{Lu2022small,Shi2021am,Volokitin2017Dey,PhysRevB.95.245437,ZHANG2022123076,10.1063/1.5132995,luo2023effect} modulation, due to {graphene's} peculiar optical properties. 
{Combining the richness coming from the grating geometry and the special dielectric features of graphene, could result in novel behaviors that cannot be obtained with ordinary materials.}
{While substantial non-additive effects in CLFs has been calculated and measured in gratings made of metals and semiconductors, these effects are not tunable  {\it{in-situ}}. So far, modifying the non-additive effects requires changing the geometric configuration. The ability to tune the non-additive effects {\it{in-situ}} could open new opportunities in the exploitation of CLFs in nano-electromechanical systems.}

\begin{figure} [htbp]
\vspace{1em}
\centerline {\includegraphics[width=0.4\textwidth]{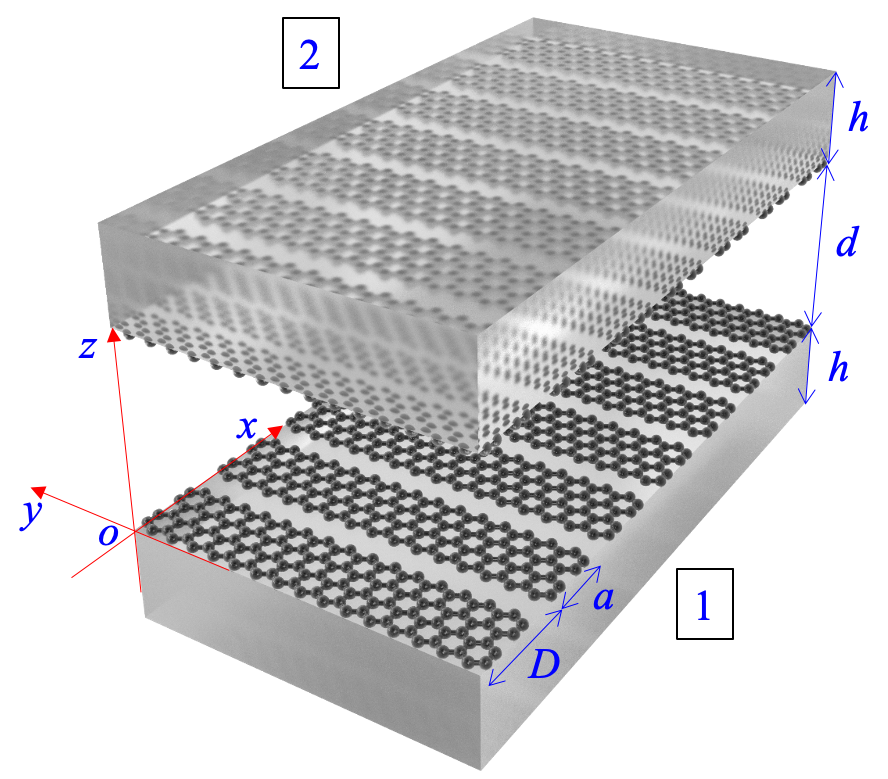}}
\caption{Schematic of two parallel graphene-gratings coated slabs. }
\label{structure}
\end{figure}

{Here we explore the CLF between two parallel graphene-based nanostructures (body 1 and body 2) separated by a distance $d$. Each structure comprises a finite} {dielectric} {substrate with thickness $h$, covered with a graphene strip grating, as depicted in Fig.~\ref{structure}. The complex nature of this grating-based system presents significant challenges, requiring considerable calculation time and extensive computational resources when using conventional methods, such as the classical Fourier Modal Method (FMM) \cite{Granet:96}. This lead to a practical impossibility to check the effective convergence and stability of the numerical outcomes with respect to the grating diffraction orders and frequency/momentum integration grid steps, with consequent qualitative and quantitative inaccurate predictions.}

{To overcome such difficulties we use an improved approach, the Fourier Modal Method with Local Basis Functions (FMM-LBF) \cite{Hwang2020,Youssef2023pre}, allowing for an efficient and accurate resolution of the scattering problem, fully taking into account the high-order diffractions. This allows us to study in detail how the CLF of the global system changes with the chemical potential and also how it is different  from the sum of the interactions between the elements constituting the global system, i.e. its non-additivity. The main prediction of this paper is that in this system the non-additvity can be significantly modulated {\it{in-situ}} by adjusting the graphene chemical potential, without altering the system geometry.}

In section \ref{sec:Phys_sys} we describe the physical system, in section \ref{sMethod} we introduce the FMM-LBF method and show how to build the scattering matrix. Finally, in section \ref{results} we present and comment the numerical results.

\section{\label{sec:Phys_sys}Physical system }

The graphene gratings have a period $D$ {of 1 $\mu$m}, a filling fraction $f=a/D$ and a chemical potential  $\mu${. They lay} on two {finite-size} fused silica (SiO$_2$) slabs of the same thickness $h$. The separation between the two bodies is $d$ and the whole system is at temperature $T=300$ K.  The Casimir-Lifshitz pressure (CLP) expression for this system can be explicitly given in terms of the reflection operators $\mathcal{R}^{(1)+}$ and $\mathcal{R}^{(2)-}$ of the two nanostructured bodies in the (TE, TM) basis. The pressure acting on body 1 along the positive z direction is given by \cite{Noto2014pra_dielectirc,PhysRevA.84.042102}
\begin{equation}
P(d,T,\mu)=\frac{ k_{\rm B} T}{4 \pi^2} \sum_{m=0}^{+\infty}{}'  \int_{-\frac{\pi}{D}}^{\frac{\pi}{D}} {\rm d}k_{x}  \int_{-\infty}^{+\infty} {\rm d} k_y   {\rm Tr}\left( \gamma '  \mathcal{M}  \right),
\label{CLF}
\end{equation}
with 
\begin{equation}
\mathcal{M}= (U^{(12)}\mathcal{R}^{(1)+} \mathcal{R}^{(2)-}+U^{(21)}\mathcal{R}^{(2)-} \mathcal{R}^{(1)+}).
\label{M_operator}
\end{equation}
The sum in {Eq.} \eqref{CLF} is carried out over the imaginary frequencies using the Matsubara frequencies $\xi_m=2\pi m k_{\rm B} T/ \hbar$ (the prime on the sum means that the $m=0$ term is to be divided by 2). {$k_{\rm{B}}$ is the Boltzmann constant, $\hbar$ is the reduced Planck's constant.} Here  $\gamma'=({\rm diag}(k_{zn}'),{\rm diag}(k_{zn}'))$, $k_{zn}'=\sqrt{\xi_m^2/c^2+{\bf{k}}_{n}^2}$, ${\bf{k}}_{n}=(k_{x, n},k_y)$, $k_{x, n}=k_x+n \frac{2\pi}{D}$, $k_{x}$ is in the first Brillouin zone $(-\frac{\pi}{D},\frac{\pi}{D})$, $k_y$ is in $(-\infty,\infty)$. The multiple scattering operators are given by
\begin{equation}
\begin{aligned}
U^{(12)}&=(1-\mathcal{R}^{(1)+}\mathcal{R}^{(2)-})^{-1},\\
U^{(21)}&=(1-\mathcal{R}^{(2)-}\mathcal{R}^{(1)+})^{-1}. 
\end{aligned}
\label{operators}
\end{equation}


To calculate the reflection operators of the two bodies, that contain periodic gratings, one can use the simplified version of FMM suited for surface gratings  where the fields are expanded in generalized Fourier series (the so called Rayleigh expansion) in the different homogeneous media while the periodic conductivity is expanded into its Fourier series. {Incorporating} all this into the boundary conditions yields an algebraic system linking the amplitudes of the fields in the different media. {The} latter can be recast into a form giving directly the S-matrix of the structure from which one can readily extract the reflection coefficients. However, it is important to note that  this method encounters convergence issues when dealing with TM polarization due to the singular nature of the electric field at the edges of the graphene sheets. To address this limitation, the FMM-LBF can be employed (more details in the next section). It incorporates locally defined basis functions that are specifically designed to satisfy the boundary conditions  \cite{Hwang2020,Youssef2023pre}.

Another method commonly used in CLP and heat transfer calculations is the FMM with Adaptive Spatial Resolution (FMM-ASR). This method has been specifically developed to address the challenging case of metallic gratings  \cite{Noto2014pra_dielectirc, Messina2017prb, Messina2015pra}. It {involves} a change of coordinates according to the periodicity direction of the grating ($x$-axis), which  improves the convergence process leading to much faster computations compared to the FMM, drastically reducing the computational time. The counterpart of this gain is that the 2D graphene grating has to be modelled with a finite thickness.

Based on our numerical analysis, we have found that the FMM-ASR method is primarily advantageous at very low frequencies compared to the FMM-LBF. As a result, we employ the FMM-ASR method specifically for computing only the first term of the Matsubara sum in Eq. \eqref{CLF}, while the subsequent terms are much more efficiently calculated using the FMM-LBF.

\begin{figure} [htbp]
\vspace{1em}
\centerline {\includegraphics[width=0.5\textwidth]{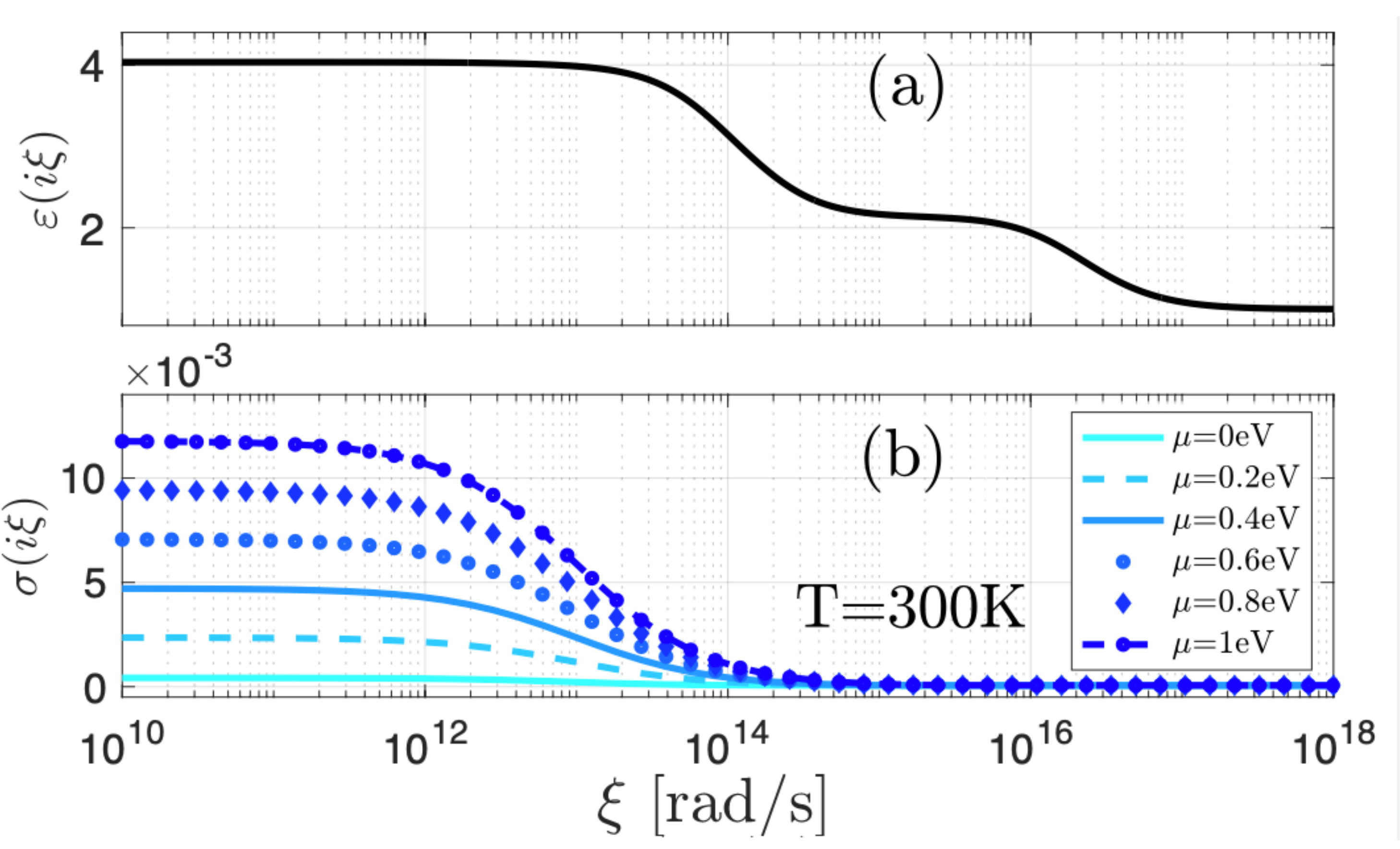}}
\caption{{(a) Relative dielectric permittivity of fused silica (Si$O_2$) and (b) graphene conductivity at $T$=300 K and for different values of $\mu$ for imaginary frequencies.}}
\label{sigma_epsilon}
\end{figure}

The dielectric function for imaginary frequencies, depicted in Fig. \ref{sigma_epsilon}(a), is derived using the Kramers-Kronig relation: $\varepsilon(i\xi_m)=1+2\pi^{-1}\int_{0}^{\infty} \omega\varepsilon''(\omega)/(\omega^2+\xi_m^2)d\omega$
and it is based on the fused silica dielectric function data along the real frequency axis obtained from [\onlinecite{Book_SiO2}].

The graphene enters through its conductivity explicitly depending on temperature $T$ and chemical potential $\mu$. It is the sum of an intraband and an interband contributions $\sigma = \sigma_{\textnormal{intra}} +\sigma_{\textnormal{inter}}$, and on  the imaginary frequency axis it takes the form [\onlinecite{Chahine2017prl},\onlinecite{Falkovsky_Graphene_1},\onlinecite{Falkovsky_Graphene_2},\onlinecite{Awan_Graphene_3}] 

\begin{equation}
\begin{aligned} 
\sigma_{\textnormal{intra}} (i\xi_m) &= \dfrac{8\sigma_0 k_B T}{\pi(\hbar \xi_m+\hbar/\tau)}\ln\left[ 2 \cosh\left(\dfrac{\mu}{2k_BT}\right) \right],\\ 
\sigma_{\textnormal{inter}}  (i\xi_m)&= \dfrac{\sigma_0 4\hbar \xi_m}{\pi}\int_0^{+\infty} \dfrac{G(x)}{(\hbar \xi_m)^2 + 4x^2} dx,
\label{sig_tot}
\end{aligned}
\end{equation} 

where, $\sigma_0={e^2}/({4\hbar})$, $e$  is the electron charge, $G(x) = \sinh (x/k_B T) / [\cosh (\mu/k_B T) + \cosh(x/k_B T)]$, $\tau$ the relaxation time (we use $\tau = 10^{-13}${s}). Fig. \ref{sigma_epsilon}(b) gives the graphene conductivity for various values of the chemical potential at $T$=300 K.\\

\begin{figure} [ht]
\centering
\includegraphics[scale=0.32]{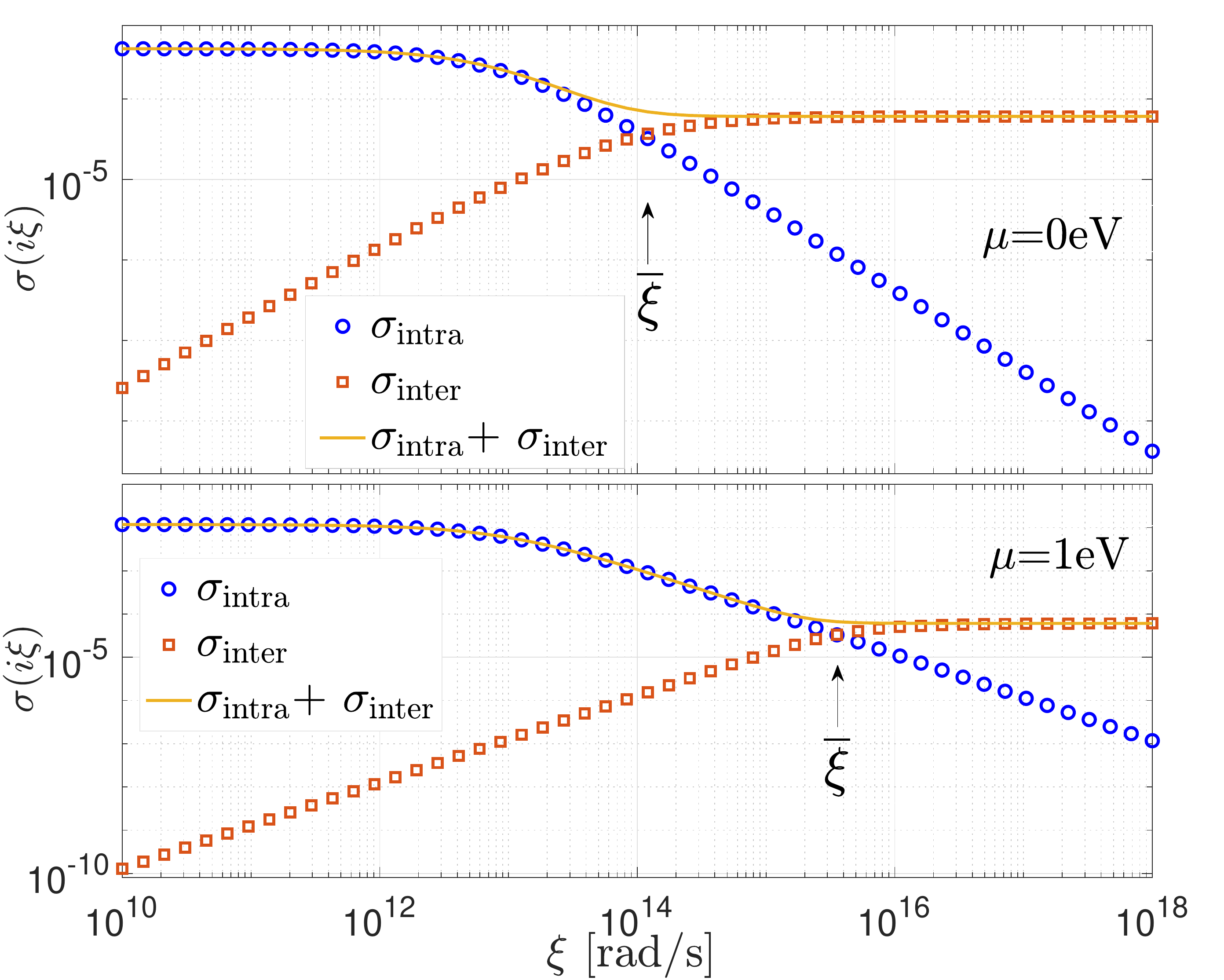}
\caption{Graphene conductivity for $\mu$=0 eV (a) and $\mu$=1 eV (a) at $T$=300 K for imaginary frequencies.}
\label{fig:inter_intra}
\end{figure}

In Fig \eqref{fig:inter_intra}, we plot each contribution as well as the total conductivity for $\mu$ = 0 eV (Fig \eqref{fig:inter_intra}(a)) and $\mu$ = 1 eV  (Fig \eqref{fig:inter_intra}(b)) at $T$=300 K. {We can clearly see that the intraband and interband contributions prevail at low and high frequencies, respectively.
 The two terms have an equivalent contribution to the total conductivity at some frequency $\overline{\xi}$ : for ${\xi} << \overline{\xi}$ the intraband prevails and for ${\xi} >> \overline{\xi}$ the interband prevails. More precisely,  $\overline{\xi}$ =$10^{14}$rad/s for $\mu$ = 0 eV and  $\overline{\xi}$ = 3 $\times$ $10^{15}$ rad/s for $\mu$ = 1 eV.}

\section{\label{sMethod}Method : The Fourier Modal Method with local basis functions}
\begin{figure} [ht]
\includegraphics[scale=0.46]{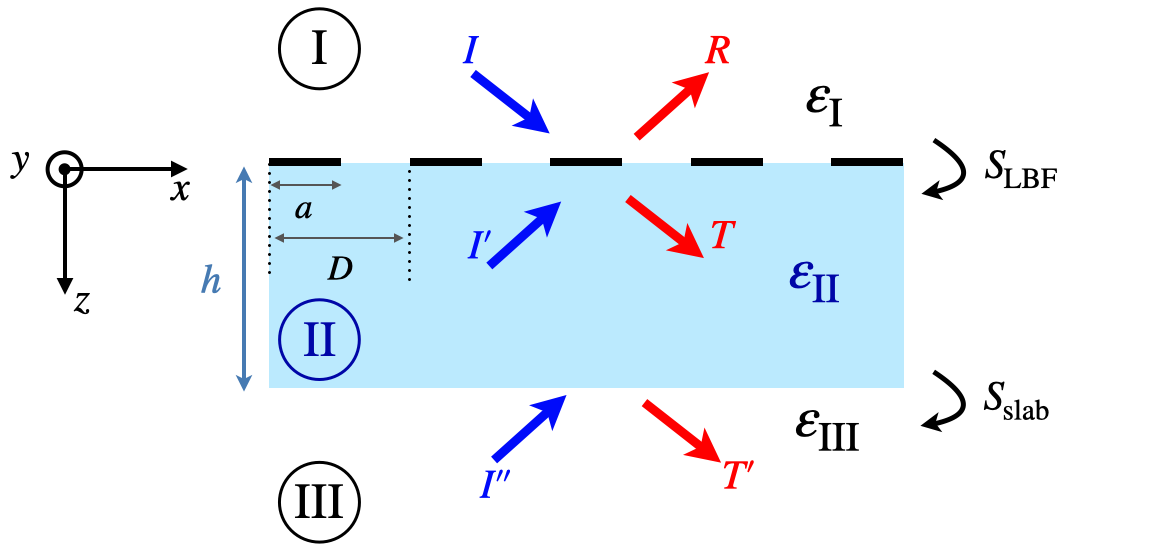}
\caption{Schematic representation of the system under study. The object consists of a finite fused silica slab with a thickness $h$ covered by a graphene grating with a period $D$, width $a$ (filling fraction is defined as $f=a/D$) and surface conductivity $\sigma$. }
\label{fig:configuration}
\end{figure}

In this section, we provide the calculation of the reflection matrices for a finite slab with a thickness $h$ and covered with a graphene strips grating characterized by a period $D$, a width $a$ and a surface conductivity $  \sigma$, as shown in Fig. $\ref{fig:configuration}$. 

The calculation employs the S-matrix algorithm, where we first compute the interface scattering matrix, denoted $S_{\rm LBF}$, between the input medium $\rm I$ and medium $\rm II$, using the FMM-LBF. Subsequently, we determine the slab scattering matrix, denoted $S_{\rm slab}$, between medium $\rm II$ and medium $\rm III$. Notably, the calculation of $S_{\rm LBF}$ and $S_{\rm slab}$is applicable in general cases; however, for the overall S-matrix to be valid, we specifically consider vacuum as both the entry and exit medium for $S_{\rm LBF}$, and vacuum as the output medium for $S_{\rm slab}$. By performing the star product ($\star$) operation (see Eq. \eqref{star_prod}) between $S_{\rm LBF}$ and $S_{\rm slab}$, we obtain the overall scattering matrix, denoted $S$, as shown in Eq. \eqref{star_product_S}. It is worth mentioning that this calculation also accommodates imaginary Matsubara frequencies {by setting $\omega=i\xi_n$}.
\begin{equation}
S = S_{\rm LBF} \star S_{\rm slab}.
\label{star_product_S}
\end{equation}

Let us begin by the first interface scattering matrice $S_{\rm LBF}$.

\subsection{Electromagnetic Fields}

Due to the periodicity along the $x$ direction, new diffraction channels open up, which can be characterized by the wave vector component along that direction. {The $z$-component of the n$^{\rm th}$ diffraction order wave vector} is dependent on the media. $ k_{zn}^{\rm I}$ and $ k_{zn}^{\rm II}$ are the $z$ wave vector of $n^{\rm th}$ order diffraction for medium I ($\varepsilon_{\rm I}$ incidence side) and medium $\rm II$ ($\varepsilon_{\rm II}$, output side), respectively and given by
\begin{equation}
\left\{
\begin{aligned}
k_{zn}^{\rm I}=\sqrt{k_0^2\varepsilon_{\rm I}- k_{xn}^2- k_y^2}, \\
k_{zn}^{\rm II}=\sqrt{k_0^2\varepsilon_{\rm II}- k_{xn}^2- k_y^2} ,
\end{aligned}
\right.
\label{gamma_I_II}
\end{equation}
with $k_0 = \omega/c$.

The electric field in medium I can be expressed as
\begin{equation}
\textbf{E}_{\rm I}=\sum_{n} (\textbf{I}_n e^{i{\textbf{k}}_{{\rm i}n}\cdot \textbf{r}}+\textbf{R}_n e^{i{\textbf{k}}_{{\rm r}n}\cdot \textbf{r}}),
\label{Conical_E_field_I}
\end{equation}
where $\textbf{I}_n=(I_{xn},I_{yn},I_{zn})$, $\textbf{R}_n=(R_{xn},R_{yn},R_{zn})$, ${\textbf{k}}_{{\rm i}n}=( k_{xn}, k_y, k_{zn}^{\rm I})$ and ${\textbf{k}}_{{\rm r}n}=( k_{xn}, k_y,- k_{zn}^{\rm I})$ and $n \in \mathbb{Z}$. Usually, in the numerical implementation, we will keep only $2N+1$ Fourier coefficients i.e. $n \in [-N,N]$; where $N$ is called the truncation order.

Then the magnetic field in medium I is
\begin{equation}
\textbf{H}_{\rm I}=\frac{1}{k_0 Z_0}\sum_n ({\textbf{k}}_{{\rm i}n} \times \textbf{I}_n e^{i{\textbf{k}}_{{\rm i}n}\cdot \textbf{r}}+{\textbf{k}}_{{\rm r}n} \times \textbf{R}_n e^{i{\textbf{k}}_{{\rm r}n}\cdot \textbf{r}}),
\label{Conical_H_field_I}
\end{equation}
where  $Z_0=\sqrt{\dfrac{\mu_0}{\varepsilon_0}}$.

The electric field in medium II is
\begin{equation}
\textbf{E}_{\rm II}=\sum_n (\textbf{T}_n e^{i{\textbf{k}}_{{\rm t}n}\cdot \textbf{r}}+\textbf{I}_n' e^{i{\textbf{k}}_{{\rm i'}n}\cdot \textbf{r}}),
\label{Conical_E_field_II}
\end{equation}
where $\textbf{T}_n=(T_{xn},T_{yn},T_{zn})$, $\textbf{I}_n'=(I_{xn}',I_{yn}',I_{zn}')$, ${\textbf{k}}_{{\rm t}n}=( k_{xn}, k_y, k_{zn}^{\rm II})$, ${\textbf{k}}_{{\rm i'}n}=( k_{xn}, k_y,- k_{zn}^{\rm II})$. Then the magnetic field in medium II is
\begin{equation}
\textbf{H}_{\rm II}=\frac{1}{k_0 Z_0}\sum_n ({\textbf{k}}_{{\rm t}n} \times \textbf{T}_n e^{i{\textbf{k}}_{{\rm t}n}\cdot \textbf{r}}+{\textbf{k}}_{{\rm i'}n} \times \textbf{I}_n' e^{i{\textbf{k}}_{{\rm i'}n}\cdot \textbf{r}}),
\label{Conical_H_field_II}
\end{equation}

In addition, using ${\rm div}\textbf{E}=\textbf{k} \cdot \textbf{E}=0$, we have the following relations
\begin{equation}
\begin{aligned}
& I_{zn}=-\frac{1}{ k_{zn}^{\rm I}}( k_{xn} I_{xn}+ k_yI_{yn}), \\
&R_{zn}=\frac{1}{ k_{zn}^{\rm I}}( k_{xn} R_{xn}+ k_y R_{yn}),
\\& T_{zn}=-\frac{1}{ k_{zn}^{\rm II}}( k_{xn} T_{xn}+ k_yT_{yn}), \\
&I_{zn}'=\frac{1}{ k_{zn}^{\rm II}}( k_{xn} I_{xn}'+ k_y I_{yn}').
\label{Iz_Rz_Tz_Izprime}
\end{aligned}
\end{equation}

\subsection{Boundary conditions}
The boundary conditions for the electric field at $z=0$ are
\begin{equation}
\left\{
\begin{aligned}
E_{{\rm I}x}(x,y,0)=E_{{\rm II}x}(x,y,0),\\
E_{{\rm I}y}(x,y,0)=E_{{\rm II}y}(x,y,0).
\end{aligned}
\right.
\label{conical_Exy_boundary}
\end{equation}

By inserting Eqs. (\ref{Conical_E_field_I}) and (\ref{Conical_E_field_II}) into Eq. (\ref{conical_Exy_boundary}),  for arbitrary $n$, we have

\begin{equation}
\left\{
\begin{aligned}
I_{xn}+R_{xn}=I_{xn}'+T_{xn},\\
I_{yn}+R_{yn}=I_{yn}'+T_{yn},
\end{aligned}
\right.
\label{E_n}
\end{equation}
which can be written in compact form
\begin{equation}
I+R=I'+T,
\label{conical_E_boundary}
\end{equation}
where 
\begin{equation}
I=\begin{pmatrix}
I_x\\
I_y
\end{pmatrix},R=\begin{pmatrix}
R_x\\
R_y
\end{pmatrix},I'=\begin{pmatrix}
I_x'\\
I_y'
\end{pmatrix},T=\begin{pmatrix}
T_x\\
T_y
\end{pmatrix},
\label{Ix_Iy}
\end{equation}

as depicted in Fig. \eqref{fig:configuration}.

Due to the zero thickness approximation of the graphene grating, the boundary condition for the magnetic fields at the interface between media I and II are
\begin{equation}
H_{{\rm II}x}(x,y,0)-H_{{\rm I}x}(x,y,0)=\sigma(x) E_{{\rm II}y}(x,y,0),
\label{1st_BDC}
\end{equation}
where the function $\sigma(x)$ is periodic and can be expanded into Fourier series as follows
\begin{equation}
\sigma(x) = \begin{cases}
  \sigma &\text{ if  }  \;\;0 \leq x < a \\
0 &\text{ if }  \;\;a \leq x < D 
\end{cases} = \sum_{n'} \sigma_{n'} e^{i\frac{2\pi}{D}n'x},
\end{equation}
where $\sigma$ comes from Eq. \eqref{sig_tot}.

It is worth noting that both $E_{{\rm I}y}(x,y,0)$ and $E_{{\rm II}y}(x,y,0)$ can be used to obtain the scattering matrices (because  $E_{{\rm (I/II)}y}(x,y,0)$ is continuous), here, we use $E_{{\rm II}y}(x,y,0)$. Then By inserting Eqs. (\ref{Conical_H_field_I}), (\ref{Conical_H_field_II}) and (\ref{Conical_E_field_II}) into Eq. (\ref{1st_BDC}), and using the Laurent factorization rule, the following relation is obtained
\begin{equation}
\begin{aligned}
& \left\{( k_y T_{zn}- k_{zn}^{\rm II} T_{yn})+( k_y I_{zn}'+ k_{zn}^{\rm II} I_{yn}')\right.\\
&\left.-( k_y I_{zn}- k_{zn}^{\rm I} I_{yn})-( k_y R_{zn}+ k_{zn}^{\rm I} R_{yn})\right\}\\
&=k_0 Z_0 \sum_{n'}\left\{\sigma_{n'-n} (T_{yn'}+I_{yn'}')\right\}.
\end{aligned}
\label{1_Hx_boundary_n}
\end{equation}

That can be expressed in a compact matrix form as

\begin{equation}
\begin{aligned}
&( k_y T_{z}-\gamma_{\rm II} T_{y})+( k_y I_{z}'+\gamma_{\rm II} I_{y}')-( k_y I_{z}-\gamma_{\rm I} I_{y})\\
&-( k_y R_{z}+\gamma_{\rm I} R_{y})=k_0 Z_0 [[\sigma]] (T_{y}+I_{y}'),
\end{aligned}
\label{1_Hx_boundary_matrix}
\end{equation}
where $\gamma_{\rm II}=$diag($k_{zn}^{\rm II}$), $\gamma_{\rm I}=$diag($k_{zn}^{\rm I}$) and $[[\sigma]]$ denotes the Toeplitz matrix whose ($n$', $n$) element is $ \sigma_{n'-n} $, more specifically:
\begin{equation}
\begin{aligned}
[[\sigma]]=
\begin{pmatrix}
\sigma_0 & \sigma_{-1} & & & & \sigma_{-2N}\\
\sigma_{1} & \sigma_0 & \sigma_{-1} & &  &\\
& \ddots & \ddots & \ddots &\\ & & \ddots & \ddots & \sigma_{-1} &\\ \sigma_{2N}^{}& & &  \sigma_{1}& \sigma_0& 
\end{pmatrix}.
\label{topleitz_matrix}
\end{aligned}
\end{equation}

Using Eq. \eqref{Iz_Rz_Tz_Izprime}, we eliminate the $z$ components in Eq. \eqref{1_Hx_boundary_matrix} to obtain 

\begin{widetext}
\begin{equation}
\begin{aligned}
-\left( \alpha k_y \gamma_{\rm II}^{-1} T_x+(\gamma_{\rm II}+\ k_y^2  \gamma_{\rm II}^{-1})T_y\right) &+\left( \alpha k_y \gamma_{\rm II}^{-1} I'_x+(\gamma_{\rm II}+{ k_y^2}{\gamma_{\rm II}}^{-1})I'_y\right) +
\left( {\alpha k_y}{\gamma_{\rm I}}^{-1} I_x+(\gamma_{\rm I}+{ k_y^2}{\gamma_{\rm I}}^{-1})I_y\right) \\
	&+\left( {\alpha k_y}{\gamma_{\rm I}}^{-1} R_x+(\gamma_{\rm I}+{ k_y^2}{\gamma_{\rm I}}^{-1})R_y\right) 
=k_0 Z_0 [[\sigma]] (T_{y}+I_{y}'),
\label{1_Hx_boundary_matrix_f}
\end{aligned}
\end{equation}
\end{widetext}
%
where $\alpha=$diag($k_{xn}$).

On the other hand, and according to \cite{Hwang2020}, the electric field ($E_x$) on the graphene grating surface ($z=0$) can be expressed in terms of the local basis functions ($g_m(x)$ and $s_m(x)$ given in Eq. \eqref{LBFs}) as
\begin{equation}
E_{x}(x,y)=e^{i k_yy}\left\{
\begin{aligned}
&\sum_{m=1}^{N_g}p_m g_m(x) \;\;\;\;\;\text{for}\;\;\; x\in\text{ graphene}\\
&\sum_{m=0}^{N_s-1}q_m s_m(x) \;\;\;\;\;\text{for}\;\;\; x\in\text{ slit}\\
\end{aligned}
\right.,
\label{E_x_FMM_LBF}
\end{equation}
where 
\begin{equation}
\;\;\;\;\;\;\;\;\;\;\;\;\;\;\;\;\;\;\;\;\;\;\;\left\{
\begin{aligned}
&g_m(x) = \sin(m \pi x/a) \\\\
&s_m(x) = \dfrac{\cos(m\pi(x-a)/c')}{\sqrt{(c'/2)^2-(x-x_c)^2}} 
\end{aligned}
\right.,
\label{LBFs}
\end{equation}
with $c'=D-a$, $x_c = (a+D)/2$, $N_g = round\left[ \dfrac{N\times a}{D}\right]$ and $N_s = N-N_g$, where $round$ stands for the nearest integer number. 

Now, the boundary conditions for the {$y$-component of the} magnetic field can be expressed as
\begin{equation}
H_{{\rm II}y}(x,y,0)-H_{{\rm I}y}(x,y,0)=-\sigma(x) E_{x}(x,y).
\label{conical_Hy_boundary}
\end{equation}
By substituting Eqs. (\ref{Conical_H_field_I}), (\ref{Conical_H_field_II}) and \eqref{E_x_FMM_LBF} in Eq. \eqref{conical_Hy_boundary} and projecting on $e^{-i k_{xn} x}$ we get
\begin{equation}
\begin{aligned}
( k_{zn}^{\rm II} T_{xn}-& k_{xn} T_{zn}- k_{zn}^{\rm II} I_{xn}'- k_{xn} I_{zn}')\\
&-( k_{zn}^{\rm I} I_{xn}- k_{xn} I_{zn}- k_{zn}^{\rm I} R_{xn}- k_{xn} R_{zn})\\
&=-  \sigma k_0 Z_0 \sum_{m=1}^{N_g}p_m G_{nm}.
\label{conical_Hy_boundary_n_single_2}
\end{aligned}
\end{equation}
%
%
%
%
%
%
where 
\begin{equation}
\begin{aligned}
G_{nm}=\frac{1}{D}\int_0^a  g_m(x) \; &e^{-i k_{xn} x} {\rm d}x \\
= \frac{-ia}{2D} e^{-i k_{xn} a/2} &\; \left[ \right.  e^{im\pi/2} sinc(\alpha_{nm}^- a/2)  \\
&-e^{-im\pi/2}  sinc(\alpha_{nm}^+ a/2) \left.\right],
\end{aligned}
\end{equation}
 and $\alpha_{nm}^\pm = m\pi/a \pm  k_{xn}$.\\

Eq. (\ref{conical_Hy_boundary_n_single_2}) can now be recast in a more compact form 
\begin{equation}
\begin{aligned}
\gamma_{\rm II} T_{x}-\alpha T_{z}-\gamma_{\rm II} I_{x}'-\alpha I_{z}'&-\gamma_{\rm I} I_{x}+\alpha I_{z}+\gamma_{\rm I} R_{x}+\alpha R_{z}\\
=&-  \sigma k_0 Z_0 \mathbb{G} p,
\label{conical_Hy_boundary_matrix1}
\end{aligned}
\end{equation}
where $\mathbb{G}=\{G_{nm}\}$ is a matrix with size ($(2N+1) \times N_g$), $p$ is the column vector formed by the $N_g$ coefficients $p_m$ {\it i.e. }$p=(p_1,p_2,p_3 \cdots p_{N_g})^{\rm T}$ and  $q$ is the column vector formed by the $N_s$ coefficients $q_m$ {\it i.e.} $q=(q_0,q_1,q_2 \cdots q_{N_s-1})^{\rm T}$.

We can write the above equation in a more suitable form
\begin{equation}
\begin{aligned}
\gamma_{\rm II} T_{x}-\alpha T_{z}-\gamma_{\rm II} I_{x}'-&\alpha I_{z}'-\gamma_{1} I_{x}+\alpha I_{z}+\gamma_{1} R_{x}+\alpha R_{z}\\
=&-  \sigma k_0 Z_0 \left[ \mathbb{G} \; \textbf{0} \right] \begin{pmatrix} p\\ q\end{pmatrix},
\label{conical_Hy_boundary_matrix1_a_b}
\end{aligned}
\end{equation}
%
where $\left[ \mathbb{G} \; \textbf{0} \right]$ is the  horizontal concatenation of matrices $\mathbb{G}$ and the matrix \textbf{0} denoting the zero matrix of size ($(2N+1) \times N_s$).

To obtain $\mathbb{G} p$, we need to take advantage of the $x$ component electric field boundary condition in the following way
\begin{equation}
E_{{\rm II}x}(x,y,0)=E_{x}(x,y).
\label{conical_Ex_boundary}
\end{equation}

By substituting Eqs. (\ref{Conical_E_field_II}) and (\ref{E_x_FMM_LBF}) into Eq. (\ref{conical_Ex_boundary}), we get the following condition
\begin{equation}
\begin{aligned}
\sum_n (T_{xn}+ I_{xn}')e^{i k_{xn} x}= \left\{
\begin{aligned}
\sum_{m=1}^{N_g}p_m g_m(x)\\
\sum_{m=0}^{N_s-1}q_m s_m(x)
\end{aligned}
\right..
\label{conical_Ex_boundary_LBF_n}
\end{aligned}
\end{equation}

By following the same procedure as before, we obtain

\begin{equation}
\begin{aligned}
&T_{xn}+ I_{xn}' 
= \sum_{m=1}^{N_g}p_m G_{nm}+\sum_{m=0}^{N_s-1}q_m S_{nm},
\label{conical_Ex_boundary_LBF_single_n}
\end{aligned}
\end{equation}
%
where 
\begin{equation}
\begin{aligned}
S_{nm}&= \frac{1}{D}\int_a^D  s_m(x)  e^{-i k_{xn} x} {\rm d}x \\
&= \frac{\pi}{2D}e^{-i k_{xn} x_c}\left[ \right. e^{im\pi/2} J_0(\beta_{nm}^- c'/2) \\
&+  e^{-im\pi/2} J_0(\beta_{nm}^+ c'/2) \left. \right],
\end{aligned}
\end{equation}
%
$\beta_{nm}^\pm = m\pi/c' \pm k_{xn}$ and $J_0$ is the zero order Bessel function of the first kind.

In compact matrix form Eq. (\ref{conical_Ex_boundary_LBF_single_n}) becomes
\begin{equation}
\begin{aligned}
T_{x}+ I_{x}'= \mathbb{G} p  +  \mathbb{S} q,
\label{conical_Ex_boundary_LBF_matrix2}
\end{aligned}
\end{equation}
where $\mathbb{S}=\{S_{nm}\}$ is a matrix with size ($(2N+1) \times N_s$).

We can write the above equation as follows
\begin{equation}
\begin{aligned}
\begin{pmatrix} p\\ q\end{pmatrix}=\left[ \mathbb{G} \; \mathbb{S} \right]^{-1} (T_{x}+ I_{x}'),
\label{a_b_matrix}
\end{aligned}
\end{equation}
where $\left[ \mathbb{G} \; \mathbb{S} \right]$ is the horizontal concatenation of matrices $\mathbb{G}$ and  $\mathbb{S}$.

By replacing Eq. (\ref{a_b_matrix}) into Eq. (\ref{conical_Hy_boundary_matrix1_a_b}), we have 
\begin{equation}
\begin{aligned}
\gamma_{\rm II} T_{x}-\alpha T_{z}-\gamma_{\rm II} I_{x}'-\alpha I_{z}'-\gamma_{\rm I} I_{x}+\alpha I_{z}+\gamma_{\rm I} R_{x}+\alpha R_{z} \\
=-   k_0 Z_0 \left[ \mathbb{G} \; \textbf{0} \right] \left[ \mathbb{G} \; \mathbb{S} \right]^{-1} (T_{x}+ I_{x}').
\label{conical_Hy_boundary_matrix}
\end{aligned}
\end{equation}

Which, by using Eqs. (\ref{Iz_Rz_Tz_Izprime}) and (\ref{1_Hx_boundary_matrix_f}), becomes
\begin{widetext}
\begin{equation}
\begin{aligned}
\begin{pmatrix}
\gamma_{\rm II}+{\alpha^2}{\gamma_{\rm II}}^{-1}+  \sigma k_0 Z_0 \left[ \mathbb{G} \; 0 \right] \left[ \mathbb{G} \; \mathbb{S} \right]^{-1} & {\alpha  k_y}{\gamma_{\rm II}^{-1}}\\
{\alpha  k_y}{\gamma_{\rm II}}^{-1} & \gamma_{\rm II} +{ k_y^2}{\gamma_{\rm II}}^{-1}+[[\sigma]] k_0 Z_0
\end{pmatrix} \begin{pmatrix}
T_x\\
T_y
\end{pmatrix}\\
-\begin{pmatrix}
\gamma_{\rm II}+{\alpha^2}{\gamma_{\rm II}}^{-1}-  \sigma k_0 Z_0 \left[ \mathbb{G}\; 0 \right] \left[ \mathbb{G}\; \mathbb{S} \right]^{-1} & {\alpha  k_y}{\gamma_{\rm II}}^{-1}\\
{\alpha  k_y}{\gamma_{\rm II}}^{-1} & \gamma_{\rm II} +{ k_y^2}{\gamma_{\rm II}}^{-1}-[[\sigma]] k_0 Z_0 
\end{pmatrix} \begin{pmatrix}
I_x'\\
I_y'
\end{pmatrix}
\\
=\begin{pmatrix}
\gamma_{\rm I}+{\alpha^2}{\gamma_{\rm I}}^{-1} & {\alpha  k_y}{\gamma_{\rm I}}^{-1}\\
{\alpha  k_y}{\gamma_{\rm I}}^{-1} & \gamma_{\rm I} +{ k_y^2}{\gamma_{\rm I}}^{-1}
\end{pmatrix} \begin{pmatrix}
I_x\\
I_y
\end{pmatrix}-\begin{pmatrix}
\gamma_{\rm I}+{\alpha^2}{\gamma_{\rm I}}^{-1} & {\alpha  k_y}{\gamma_{\rm I}}^{-1}\\
{\alpha  k_y}{\gamma_{\rm I}}^{-1} & \gamma_{\rm I} +{ k_y^2}{\gamma_{\rm I}}^{-1}
\end{pmatrix} \begin{pmatrix}
R_x\\
R_y
\end{pmatrix}.
\label{conical_H_boundary_detailed}
\end{aligned}
\end{equation}
\end{widetext}

Eq. (\ref{conical_H_boundary_detailed}) can be written in a more compact form
\begin{equation}
(A+\Lambda)T+(\Lambda-A)I'=B(I-R),
\label{conical_H_boundary}
\end{equation}
where $\Lambda={\rm diag(  \sigma k_0 Z_0 \left[ \mathbb{G} \; 0 \right] \left[ \mathbb{G} \;\mathbb{S} \right]^{-1},[[\sigma]] k_0 Z_0)}$, $A$ and $B$ are defined as follows
\begin{equation}
\begin{aligned}
A&=\begin{pmatrix}
\gamma_{\rm II}+{\alpha^2}{\gamma_{\rm II}}^{-1} & {\alpha  k_y}{\gamma_{\rm II}}^{-1}\\
{\alpha  k_y}{\gamma_{\rm II}}^{-1} & \gamma_{\rm II} +{ k_y^2}{\gamma_{\rm II}}^{-1}
\end{pmatrix}, \\
B&=\begin{pmatrix}
\gamma_{\rm I}+{\alpha^2}{\gamma_{\rm I}}^{-1} & {\alpha  k_y}{\gamma_{\rm I}}^{-1}\\
{\alpha  k_y}{\gamma_{\rm I}}^{-1} & \gamma_{\rm I} +{ k_y^2}{\gamma_{\rm I}}^{-1}
\end{pmatrix}.
\label{A_B_matrix}
\end{aligned}
\end{equation}

Combining Eqs. (\ref{conical_E_boundary}) and (\ref{conical_H_boundary}), we obtain 
\begin{equation}
\begin{pmatrix}
\mathbbm{1} & -\mathbbm{1}\\
B & A+\Lambda
\end{pmatrix} \begin{pmatrix}
R\\
T
\end{pmatrix}=\begin{pmatrix}
-\mathbbm{1} & \mathbbm{1}\\
B & A-\Lambda
\end{pmatrix} \begin{pmatrix}
I\\
I'
\end{pmatrix},
\label{Conical_Total_fields_bis}
\end{equation}
{where $\mathbbm{1}$ is the identity matrix of size $(2(2N+1) \times 2(2N+1))$.}

Finally the interface scattering matrix $S_{\rm LBF}$ is given by
\begin{equation}
\begin{pmatrix}
R\\
T
\end{pmatrix}=S_{\rm LBF} \begin{pmatrix}
I\\
I'
\end{pmatrix},
\label{Conical_Total_fields}
\end{equation}
where
\begin{equation}
S_{\rm LBF}=\begin{pmatrix}
\mathbbm{1} & -\mathbbm{1}\\
B & A+\Lambda
\end{pmatrix}^{-1} \begin{pmatrix}
-\mathbbm{1} & \mathbbm{1}\\
B & A-\Lambda
\end{pmatrix}.
\label{S_LBF_conical}
\end{equation}

\subsection{Scattering matrix of the finite slab covered with a graphene grating}
Now let us move on to the second scattering matrix $S_{\rm slab}$, it is given by 
\begin{equation}
S_{\rm slab} \begin{pmatrix}
T\\
I''
\end{pmatrix}= \begin{pmatrix}
I'\\
T'
\end{pmatrix},
\label{Conical_Total_fields_bis1}
\end{equation}
where
\begin{equation}
S_{\rm slab}=\begin{pmatrix}
{\Phi} & \mathbb{0}\\
\mathbb{0} & \mathbbm{1}
\end{pmatrix} \begin{pmatrix}
\mathbbm{1} & -\mathbbm{1}\\
\mathbbm{M}_1 & \mathbbm{M}_2
\end{pmatrix}^{-1} \begin{pmatrix}
-\mathbbm{1} & \mathbbm{1}\\
\mathbbm{M}_1 & \mathbbm{M}_2
\end{pmatrix} \begin{pmatrix}
{\Phi} & \mathbb{0}\\
\mathbb{0}& \mathbbm{1}
\end{pmatrix},
\label{S_matrix_slab}
\end{equation}
with $\Phi={\rm diag}\left({\rm diag}(e^{i\gamma_{\rm II}h}),{\rm diag}(e^{i\gamma_{\rm II}h})\right)$, $\mathbb{0}$ is the null matrix of size $(2(2N+1) \times 2(2N+1))$ and\\
\begin{equation}
\begin{aligned}
&\mathbbm{M}_{1} = \begin{pmatrix}{\alpha  k_y}{\gamma_{\rm II}}^{-1} & \gamma_{\rm II} +{ k_y^2}{\gamma_{\rm II}}^{-1}\\
-(\gamma_{{\rm II}}+{\alpha^2}{\gamma_{\rm II}}^{-1}) & -{\alpha  k_y}{\gamma_{\rm II}}^{-1}
\end{pmatrix},\\
&\mathbbm{M}_{2} = \begin{pmatrix}{\alpha  k_y}{\gamma_{\rm III}}^{-1} & \gamma_{\rm III} +{ k_y^2}{\gamma_{\rm III}}^{-1}\\
-(\gamma_{\rm III}+{\alpha^2}{\gamma_{\rm III}}^{-1}) & -{\alpha  k_y}{\gamma_{\rm III}}^{-1}
\end{pmatrix},
\end{aligned}
\end{equation}
where $\gamma_{\rm III}={\rm diag}(k_{zn}^{\rm III})$ and $k_{zn}^{\rm III}$ is the $z$ wave vector of $n^{\rm th}$ order diffraction for medium III and given by $k_{zn}^{\rm III} = \sqrt{k_0^2\varepsilon_{\rm III}-k_{xn}^2- k_y^2}$.

Finally, we obtain the total S-matrix (cf.  Eq. \eqref{star_product_S}) that connects the amplitudes of medium I to those of medium III as follows

\begin{equation}
\begin{pmatrix}
R\\
T'
\end{pmatrix}=(S_{\rm LBF} \star S_{\rm slab}) \begin{pmatrix}
I\\
I''
\end{pmatrix}=S_{\rm} \begin{pmatrix}
I\\
I''
\end{pmatrix}.
\label{Conical_Total_fields_bis_2}
\end{equation}

The operation $\mathbbm{A}= \mathbbm{B} \star\mathbbm{C}$ is defined as \cite{Messina2017prb}

\begin{equation}
\begin{aligned}
&\mathbbm{A}_{11} = \mathbbm{B}_{11} + \mathbbm{B}_{12}(\mathbbm{1}-\mathbbm{C}_{11}\mathbbm{B}_{22})^{-1} \mathbbm{C}_{11} \mathbbm{B}_{21},\\
&\mathbbm{A}_{12} = \mathbbm{B}_{12}(\mathbbm{1}-\mathbbm{C}_{11}\mathbbm{B}_{22})^{-1} \mathbbm{C}_{12},\\
&\mathbbm{A}_{21} = \mathbbm{C}_{21}(\mathbbm{1}-\mathbbm{B}_{22}\mathbbm{C}_{11})^{-1} \mathbbm{B}_{21},\\
&\mathbbm{A}_{22} = \mathbbm{C}_{22} + \mathbbm{C}_{21}(\mathbbm{1}-\mathbbm{B}_{22}\mathbbm{C}_{11})^{-1} \mathbbm{B}_{22}\mathbbm{C}_{12}.
\label{star_prod}
\end{aligned}
\end{equation}

The scattering matrix in Eq. (\ref{Conical_Total_fields_bis_2}) with a size of $(4(2N+1) \times 4(2N+1))$ defines the reflection and transmission matrices as follows 
\begin{equation}
S=\begin{pmatrix}
\mathcal{R}_{xyz}^{-} & \mathcal{T}_{xyz}^{-}\\
\mathcal{T}_{xyz}^{+} & \mathcal{R}_{xyz}^{+}
\end{pmatrix}.
\label{R_matrix_conical}
\end{equation}\

For completeness, the $\mathcal{T}_{xyz}^{-}$ and $\mathcal{T}_{xyz}^{+}$ transmission coefficients must be multplied by a phase factor $e^{-i \gamma_{\rm III} h}$. 

\subsection{Transformation matrices}
In the expression of the CLP of Eq. (\ref{CLF}), the scattering matrices are expressed in the standard surface optics TE and TM basis. In the following, we provide the details on how to change basis and how to explicitly derive the $\mathcal{R}^{(1)+}$ and $\mathcal{R}^{(2)-}$ entering in this equation.

To manipulate the transformation of reflection operators from the $(x, y, z)$ Cartesian basis to the (TE, TM) basis, we firstly need to define the unit vectors in (TE, TM) basis 
\begin{equation}
\begin{aligned}
& \hat{\bf{e}}_{\rm TE}^{\phi}({\bf{k}}_{n},\omega)=\frac{1}{k_n}(-k_y \hat{\bf{e}}_{x}+k_{xn}\hat{\bf{e}}_{y}),\\
& \hat{\bf{e}}_{\rm TM}^{\phi}({\bf{k}}_{n},\omega)=\frac{c}{\omega}(-k_n  \hat{\bf{e}}_{z} + \phi k_{zn} \hat{\bf{k}}_{n}),
\label{TE_TM_basis_vectors}
\end{aligned}
\end{equation}
where $\hat{\bf{e}}_{x}$, $\hat{\bf{e}}_{y}$ and $\hat{\bf{e}}_{z}$ are unit vectors in the $(x, y, z)$ Cartesian basis, ${\bf{k}}_{n}=(k_{xn},k_y)$, $\hat{\bf{k}}_{n}={\bf{k}}_{n}/k_n$, $k_{zn}=\sqrt{\omega^2/c^2-{\bf{k}}_{n}^2}$ and $\phi$ {is} the direction of propagation of the waves $(+, -)$ along the $z$-axis for the incident and the reflected fields, respectively.

In the $(x, y, z)$ Cartesian basis, the field of order $n$ is expressed as $\textbf{E}_n=E_{x,n} \hat{\bf{e}}_{x}+E_{y,n} \hat{\bf{e}}_{y} + E_{z,n} \hat{\bf{e}}_{z}$. In the (TE, TM) basis we can write $\textbf{E}_n=E_{{\rm TE},n} \hat{\bf{e}}_{\rm TE}+E_{{\rm TM},n} \hat{\bf{e}}_{\rm TM}$, which can be rearranged by {applying} the Eq. (\ref{TE_TM_basis_vectors}) as follows
\begin{equation}
\begin{aligned}
\textbf{E}_n&=(-\frac{k_y}{k_n} E_{{\rm TE},n}+\phi \frac{c}{\omega} \frac{k_{zn} k_x}{k_{n}}E_{{\rm TM},n})\hat{\bf{e}}_{x}\\
&+(\frac{k_x}{k_n} E_{{\rm TE},n}+\phi \frac{c}{\omega} \frac{k_{zn} k_y}{k_{n}}E_{{\rm TM},n})\hat{\bf{e}}_{y}\\
&-k_n \frac{c}{\omega} E_{{\rm TM},n} \hat{\bf{e}}_{z}.
\label{E_field_for_transformation}
\end{aligned}
\end{equation}

By comparing the above Eq. (\ref{E_field_for_transformation}) and $\textbf{E}_n=E_{x,n} \hat{\bf{e}}_{x}+E_{y,n} \hat{\bf{e}}_{y} + E_{z,n} \hat{\bf{e}}_{z}$, we obtain the following relation
\begin{equation}
\begin{pmatrix}
E_{x,n}\\
E_{y,n}
\end{pmatrix}=\begin{pmatrix}
-\frac{k_y}{k_n} & \phi \frac{c}{\omega} \frac{k_{zn} k_{xn}}{k_{n}}\\
\frac{k_{xn}}{k_n} & \phi \frac{c}{\omega} \frac{k_{zn} k_y}{k_{n}}
\end{pmatrix} \begin{pmatrix}
E_{{\rm TE},n}\\
E_{{\rm TM},n}
\end{pmatrix}.
\label{Transformation_matrix_n}
\end{equation}

The above relationship can be expressed in a more concise form 

\begin{equation}
\begin{pmatrix}
E_{x}\\
E_{y}
\end{pmatrix}=\mathbb{B}^{\phi} \begin{pmatrix}
E_{{\rm TE}}\\
E_{{\rm TM}}
\end{pmatrix},
\label{Field_Transformation}
\end{equation}
where the transformation matrix is 
\begin{equation}
\mathbb{B}^{\phi}=\begin{pmatrix}
-{\rm diag}(\frac{k_y}{k_n}) & {\rm diag}(\frac{c \phi k_{zn} k_{xn}}{\omega k_{n}})\\
{\rm diag}(\frac{k_{xn}}{k_n}) & {\rm diag}(\frac{c \phi k_{zn} k_y}{\omega k_{n}})
\end{pmatrix}.
\label{Transformation_matrix}
\end{equation}

By applying this transformation matrix to the reflection operator $\mathcal{R}_{xyz}^{-}$, the reflection operator in the (TE, TM) basis becomes
\begin{equation}
\mathcal{R}^{-}=(\mathbb{B}^-)^{-1}~\mathcal{R}_{xyz}^{-}~\mathbb{B}^+.
\label{Reflection_operator_TE}
\end{equation}

We stress here that, due to the change of basis, the matrix $\mathcal{R}^{-}$  as defined in Eq. \eqref{Reflection_operator_TE} is now ordered in this specific basis as follows

\begin{equation}
\mathcal{R}^{-} = \begin{pmatrix}
\mathcal{R}^{-}_{TE/TE} & \mathcal{R}^{-}_{TE/TM}\\
\mathcal{R}^{-}_{TM/TE} & \mathcal{R}^{-}_{TM/TM}
\end{pmatrix}.
\label{Rminus}
\end{equation}

The size of matrix $\mathcal{R}^{-}$ is $2(2N+1) \times 2(2N+1)$ while the size of the sub matrices is $(2N+1) \times (2N+1)$.\\

In {Eqs.}~\eqref{CLF} and \eqref{M_operator} appear the reflection operators $\mathcal{R}^{(1)+}$ and $\mathcal{R}^{(2)-}$ of the two bodies of the same material, which can  be directly obtained from the matrix $\mathcal{R}^{-}$ we just derived in Eq.~\eqref{Rminus}. In particular, due to the different $z$-axis orientation, $\mathcal{R}^{(1)+}$ is identical to $\mathcal{R}^{-}$ on the two diagonal blocks (TE/TE) and (TM/TM), while it has a sign difference for the two off-diagonal blocks (TE/TM) and (TM/TE) as described in \cite{Messina2017prb}

\begin{equation}
\mathcal{R}^{(1)+} = \begin{cases}
\mathcal{R}^{-}_{p,p} & p=p'\\
-\mathcal{R}^{-}_{p,p'} & p \neq p'.
\end{cases}
\end{equation}

When body $2$ is positioned at a distance $d$ from the origin, it results in a phase shift in the reflection operator $\mathcal{R}^{(2)-}$, similar to what is described in \cite{PhysRevA.84.042102}

\begin{equation}
\begin{aligned}
\left \langle p,{\textbf{k}},n|\mathcal{R}^{(2)-}(\omega)|p',{\textbf{k}}',n' \right\rangle  \\
= e^{i\left(k_{zn}^{}+k_{zn'}'\right) d}&	\left \langle p,{\textbf{k}},n|\mathcal{R}^{-}(\omega)|p',{\textbf{k}}',n' \right\rangle.
\label{R2}
\end{aligned}
\end{equation}

We finally are able to compute the Casimir-Lifshitz pressure in Eq. (\ref{CLF}) with these two reflection operators.

\section{\label{results}Results}
\subsection{Modulation of CLP by the chemical potential}

\begin{figure} [htbp]
  \centerline{\includegraphics[width=0.48\textwidth]{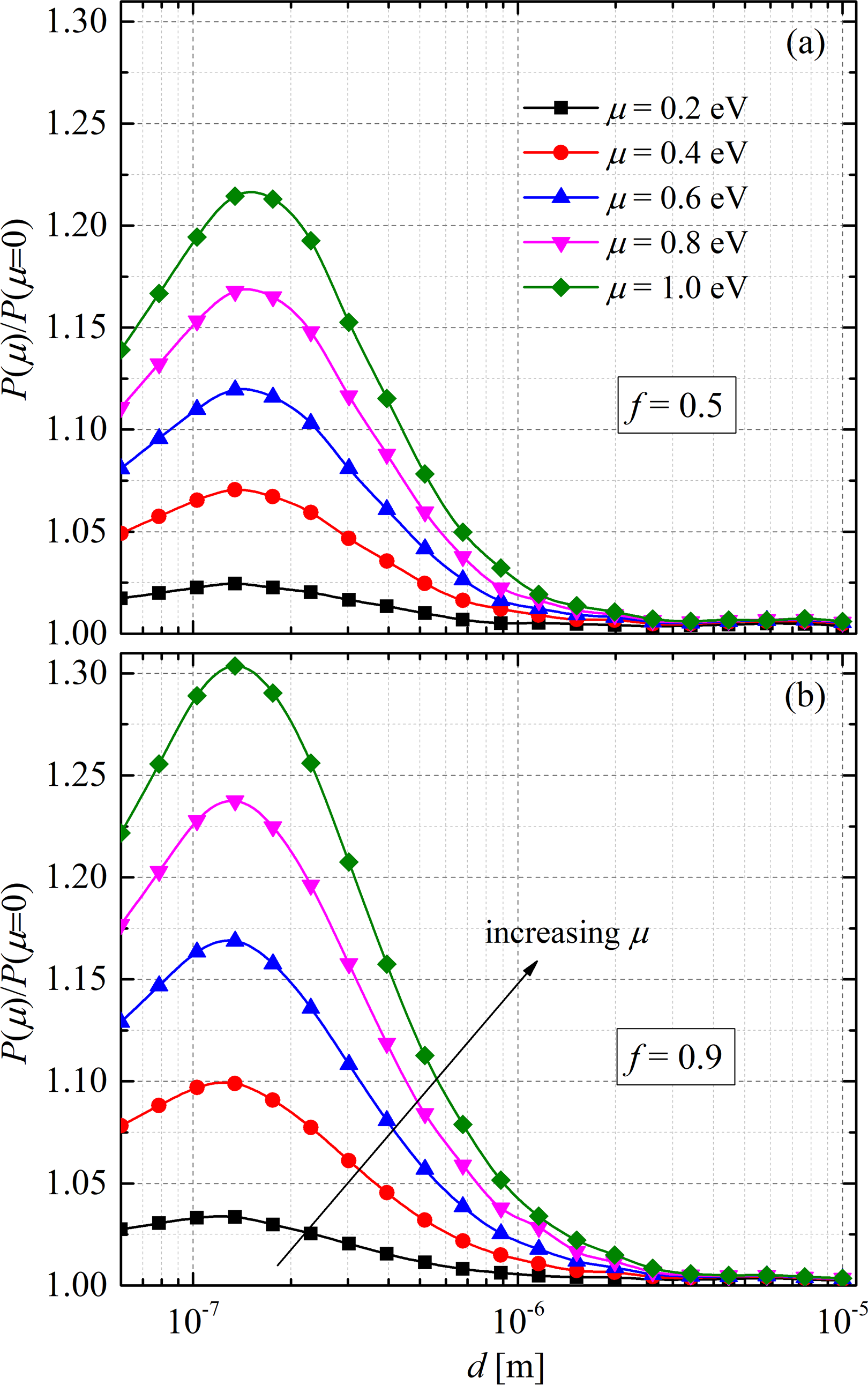}}
  \caption{Normalized CLP $P(\mu)/P(\mu=0)$ at $T$=300 K, for different chemical potentials $\mu$ and for a filling fraction (a) $f=0.5$ and (b) $f=0.9$.}
   \label{fig3:chemical_potential}
\end{figure}

We {first study} how the CLP can be tuned by changing {the} chemical potential $\mu$ of the graphene grating in our system.
We considered graphene gratings  with two different filling fractions, $f=0.5$ and 0.9, chemical potentials $\mu$ = 0, 0.2, 0.4, 0.6, 0.8, and 1.0 eV, coated on an $h=20$ nm fused silica substrate. The separation distance between the two graphene gratings varies from 60 nm to 10$~\mu$m. The dependences of the modulation ratio $ P(\mu)/P(\mu=0)$ on the separation distances $d$ are shown in Fig.~\ref{fig3:chemical_potential} (a) and (b) for $f=$ 0.5 and 0.9, respectively.

The impact of the chemical potential on the CLP of graphene gratings is found to be significant at separations less than 1 $\mu$m{. The modulation ratio attains peak values of 1.23 and 1.30 for {$f$} = 0.5 and {$f$} = 0.9 respectively and then} diminishes as separation increases {beyond 1 $\mu$m}. This trend is analogous to that observed for graphene multilayers, where the chemical potential effect was negligible at large separations \cite{Chahine2017prl}. {The modulation ratio for {$f$} = 0.9 exceeds that for {$f$} = 0.5 over the entire range of distance in our calculations.} 

\subsection{Non-additive effects}

\begin{figure} [htbp]
\centerline {\includegraphics[width=0.48\textwidth]{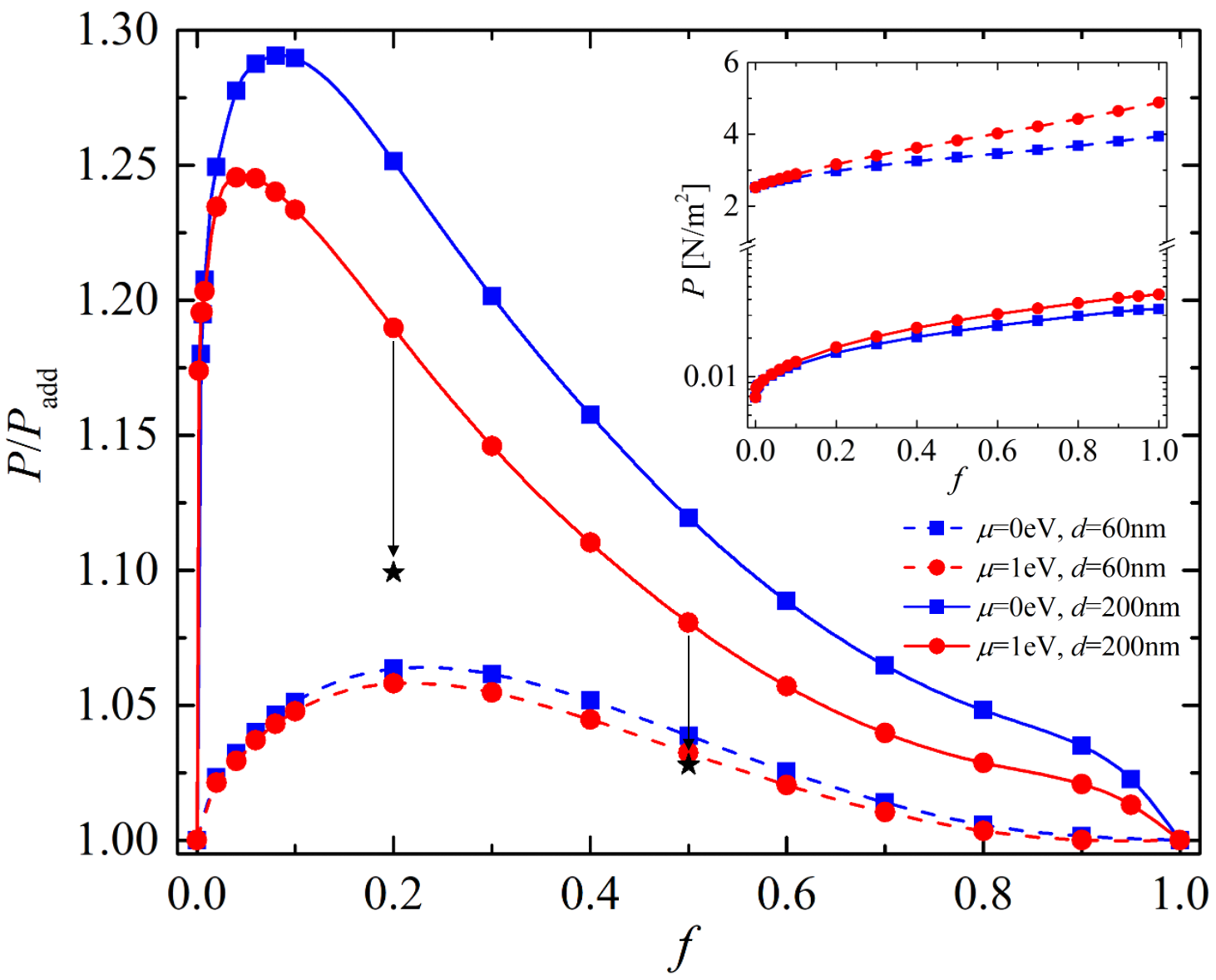}}
        \caption{Dependence of the $P/P_{\rm add}$ (main figure) and of $P$ (inset) on the filling fraction $f$ of the graphene grating.}
        \label{fig2:non_additivity}
\end{figure}

{Next, we investigate} non-additive effects in the CLP. Specifically, in Fig.~\ref{fig2:non_additivity}  we compare the full calculation coming from \eqref{CLF} which uses the complete scattering of the structured system, with {the approximation of} purely additive and much simpler calculation $P_{\rm add}=P(f=1) \times f+P(f=0) \times (1-f)${. The latter considers} the pressure as a weighed sum of the CLP occurring between planar, non-nanostructured systems, namely fully graphene-coated substrates $P(f=1)$ and graphene-free substrates $P(f=0)$. This study has been done for two main separation distances of experimental relevance ($d=60$ nm and $d=200$ nm) and also for two extreme values of the chemical potential ($\mu=0$ eV and $\mu=1$ eV). In Fig. \ref{fig2:non_additivity},  we show the ratio $P/P_{\rm add}$ for different filling fractions $f$. 
We see that the non additivity is quite weak (of the order of $5\%$) at $d=60$ nm, {with little dependence} on the chemical potential. 
This means that, at such short separations, the non-additive complexity and the particular surface mode structure of the given graphene grating have a weak effect and that one can safely use the approximate additive expression $P_{\rm add}$ for experiments with an accuracy of few percent. It is worth stressing that the calculation of the full exact CLP {is} $10^3-10^4$ slower {and much less straight-forward to be coded} than the simple additive calculation $P_{\rm add}$, {even when the faster numerical FMM-LBF method is used. Exact calculations of the CLP for graphene gratings requires} considerable computational resources and time.

On the contrary at larger but sill experimentally relevant separations, the situation changes. We see in Fig. \ref{fig2:non_additivity} that for $d=200$ nm the CLP is strongly not additive. The scattering details of the gratings are crucial, and the additive expression $P_{\rm add}$ is violated up to $30\%$. In this case a comparison with experiments needs a full theory with a complete consideration of the complexity of the nanostructure. 
 {Another} crucial point emerging from this study is that, remarkably, in this structure the non-additive effect is not only high, but can also be modulated {\it{in situ}} by simply changing the  chemical potential{. The possibility of tuning non-additive effects} without any geometric {modification is highly relevant to experimental studies}. {Specifically} for $d=200$ nm {there are clear changes in} the non-additivity $P/P_{\rm add}$ {as} the chemical potential {is increased} from 0 to 1 eV. {Substantial changes in the non-additivity occur} over a wide region of filling fraction values. Collective non-additive contributions affect the system as soon {as} a grating structuration is {introduced}, even if the strips cover only a relatively minor or major part of the substrate, and the way this non-additive effects contributes can also be easily tuned. {Similar to the case for separation of 60 nm, } the ratio $P/P_{\rm add}$ {attains} a peak and then gradually decreases as the filling fraction $f$ increases. The peak position of the $P/P_{\rm add}$ curve shifts towards lower values of the filling fraction {as the separation is increased from 60 nm to 200 nm}. 
 
  {To highlight the origin of these features, we introduce a natural dimensionless parameter $d/D$, i.e. the ratio between the separation $d$ and the grating period  $D$. While the different curves in Fig. \ref{fig2:non_additivity} correspond to the same grating period ($D=1\mu$m), they have $d/D=0.06$ for $d$=60 nm, while $d/D=0.2$ for $d$=200 nm. To investigate the relevance of the geometric factor $d/D$ we calculated $P/P_{\rm add}$ for $d$=200 nm with a larger value of $D=3.3 \mu$m, such that the ratio becomes now $d/D=0.06$. The results are shown in Fig. \ref{fig2:non_additivity} with two black stars, corresponding to filling fractions $f=0.2$ and 0.5. We see that they both move close to the $d=60$ nm curve, confirming the relevance of the geometric factor $d/D$ and hence of the geometric nature of the effect. The black star for $f=0.5$ is quite exactly on the curve corresponding to $d=60$ nm, while for $f=0.2$ it is not exactly on that curve since edge effects due to the presence of graphene strips also play a role here.}

\ {On the experimental side, the chemical potential of graphene can be controlled by chemical doping \cite{Liu} that involves charge transfer from adsorbed dopants to graphene. Alternatively, the chemical potential can also be tuned by fabricating gates underneath the graphene. The gates need to be positioned so that electric fields do not leak through the oxide regions to generate unwanted electrostatic forces.
According to Fig. \ref{fig2:non_additivity}, the CLP deviates from $P_{\rm add}$ by ~ 6$\%$ and $~$ 14$\%$ for $\mu=0$ eV and $\mu=1$ eV respectively at $d=200$ nm, $f=0.5$ and $T=300$ K. The corresponding Casimir pressures are 0.023 Nm$^{-2}$ and 0.027 Nm$^{-2}$ respectively. Assuming an effective interaction area of 10 $\mu$m by 10 $\mu$m, the force is well within the detection sensitivity of most experiments measuring CLP. However, detecting the deviation from $P_{\rm add}$ and, furthermore, distinguishing the difference in deviation for different $\mu$ would require the highest sensitivity achieved in experiments so far \cite{Bimonte}.}

\section{Conclusion}
In summary, we have studied the Casimir interactions in  graphene nanostructures made of graphene gratings coated on dielectric slabs. To fully take into consideration the high-order diffractions in the CLP acting on the gratings of the two-dimensional materials, we applied an exact method {using} the FMM-LBF. We first find a significant variation of the CLP with the chemical potential. We then studied the non-additivity and find that at small separation ($d=60$ nm) the non-additivity is weak{. As a result} the much direct and fast approximate additive method can be safely used to calculate the force at few percent precision. {We show that the geometric parameter $d/D$ is pertinent, and if one keeps it relatively small ($d/D<0.06$) the exact pressure differs from the very simple $P_{\rm add}$ by only few percent (which is typically enough for experimental comparison purposes) when $f \gtrsim  0.5$.
On the contrary, a significant non-additive effect is present at larger values of $d/D$, as we see for $d=200$ nm with $D$=1 $\mu$m, where a deviation from the additive prediction goes up to $30\%$.} Remarkably, this non additivity can be modulated {\it{in situ}} by {changing} the graphene chemical potential, without any need of geometrical or mechanical variations in the system. The presence of this non-addivity modulability {can motivate} experimental investigations of the non-additivity on CLP  {and open new opportunities of utilizing non-additive effects in graphene nano-elctromechanical systems.}

\begin{acknowledgments}
The work described in this paper was supported by a grant "CAT" from the ANR/RGC Joint Research Scheme sponsored by the French National Research Agency (ANR) and the Research
Grants Council (RGC) of the Hong Kong Special Administrative Region{, China (Project No. A-HKUST604/20)01.} We acknowledge P. Rodriguez-Lopez for useful comments.\\
Y.J. and M.L. contributed equally to this work.

\end{acknowledgments}

\section*{}
\bibliography{CLF.bib}

\end{document}